\documentclass[12pt]{article}
\usepackage{epsfig}
\usepackage[english]{babel}
\title{Near threshold radiative 3$\pi$ production in $e^+e^-$ annihilation}
\author{ A.Ahmedov$^a$, G.V.Fedotovich$^b$, E.A.Kuraev$^c$, Z.K.Silagadze$^b$
\\[.4cm]
$^a${\small Laboratory of Particle Physics,
     JINR, 141980, Dubna, 141980 Russia }\\
$^b${\small Budker Institute of Nuclear Physics, 630 090, Novosibirsk,
Russia   } \\
$^c${\small Laboratory of Theoretical Physics, JINR, 141980, Dubna,
Russia }
}
\date{}

\begin{document}
\large
\maketitle

\begin{abstract}
We consider the $\pi^+\pi^-\pi_0\gamma$ final state in electron-positron
annihilation at cms energies not far from the threshold.
Both initial and final state radiations of the hard photon are considered,
but without interference between them. The amplitude for the final state
radiation is obtained by using the effective Wess-Zumino-Witten Lagrangian
for pion-photon interactions valid for low energies. In real experiments
energies are never so small that $\rho$ and $\omega$ mesons would have a
negligible effect. So a phenomenological Breit-Wigner factor is introduced
in the final state radiation amplitude to account for the vector mesons
influence. Using radiative 3$\pi$ production amplitudes, a Monte Carlo event
generator is developed which could be useful in experimental studies.
\end{abstract}

\newpage

\section{Introduction}
The new Brookhaven experimental result for the anomalous magnetic moment
of the muon \cite{1} aroused considerable interest in the physics community,
because it was interpreted as indicating a new physics beyond the
Standard Model \cite{2}. However, such claims, too premature in our opinion,
assume that the theoretical prediction for the muon anomaly is well
understood at the level of necessary precision. Hadronic uncertainties
become of main concern \cite{3}. Fortunately the leading hadronic contribution
is related to the hadronic corrections to the photon vacuum polarization
function, which can be accurately calculated provided that the precise
experimental data on the low-energy hadronic cross sections in the $e^+e^-$
annihilation are at our disposal.

In the last few years high-statistics experimental data were collected in the
$\rho$-$\omega$ region in Novosibirsk experiments at the VEPP-2M collider
\cite{4}. In this region the hadronic cross sections are dominated by the
$e^+e^-\to 2\pi$ and $e^+e^-\to 3\pi$ channels. The former is of uppermost
importance for reduction of errors in the evaluation of the hadronic vacuum
polarization contribution to the muon g-2. Considerable progress was reported
for this channel by the CMD-2 collaboration \cite{5}. The $e^+e^-\to 3\pi$
channel, which gives a less important but still significant contribution
to the hadronic error, was also investigated in the same experiment in the
$\omega$-meson region \cite{6}. Such high precision experiments require
accurate knowledge of various backgrounds. Among them the 
$e^+e^-\to 3\pi\gamma$
channel provides an important background needed to be well understood. This
experimental necessity motivated our investigation of the three pion radiative
production presented here. Besides, being of interest as an important
background source, this process could be of interest by itself, because
a detailed experimental study of the final state radiation will allow one to 
get important information about pion-photon dynamics at low energies. 
However, such an
experimental investigation will require much more statistics than available
in VEPP-2M experiments and maybe would be feasible only at $\phi$-factories
where the low energy region can be reached by radiative return technique
as was recently been demonstrated in the KLOE experiment \cite{7}.

\section{Initial state radiation}
\begin{figure}[htb]
  \begin{center}
\mbox{\epsfig{figure=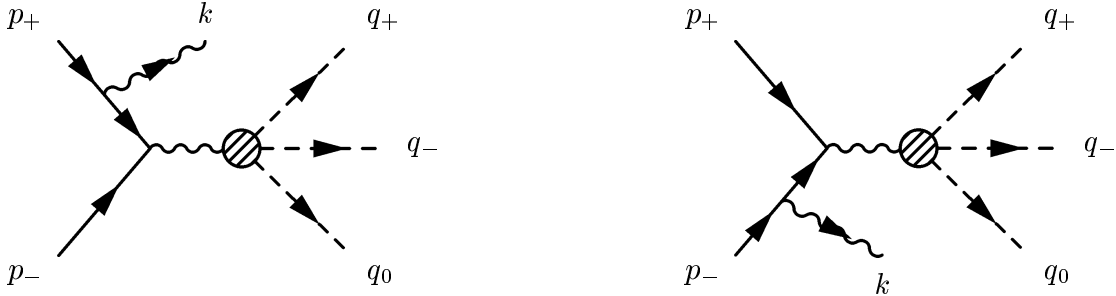
                               ,  height=4.0cm}}
   \end{center}
\caption {Initial state radiation diagrams and particle 4-momenta
assignment.}
\label{Fig1}
\end{figure}

Let $J_\mu$ be the matrix element of the electromagnetic current between the
vacuum and the $\pi^+\pi^-\pi^0$ final state. Then the initial state
radiation (ISR) contribution to the $e^+e^-\to\pi^+\pi^-\pi^0\gamma$ process
cross section is given at $O(\alpha^3)$ by the standard expression \cite{8}
\begin{eqnarray} & & d\sigma_{ISR}(e^+e^-\to 3\pi\gamma)=\frac{e^6}{4(2\pi)^8
(Q^2)^2}\left \{ \frac{Q^2}{4E^2}~J\cdot J^* \left (\frac{p_+}{k\cdot p_+}-
\frac{p_-}{k\cdot p_-} \right )^2- \right . \nonumber \\ & & \left .
-\frac{Q^2}{2E^2}~\frac{p_+\cdot J~
p_+\cdot J^* + p_-\cdot J~ p_-\cdot J^*}{k\cdot p_+ k\cdot p_-}-
\frac{J\cdot J^*}{2E^2}\left (
\frac{k\cdot p_+}{k\cdot p_-}+\frac{k\cdot p_-}{k\cdot p_+}\right )
+ \right .  \label{eq1} \\ & & \left . +\frac{m_e^2}{E^2}
\left ( \frac{p_+\cdot J} {k\cdot p_-}-\frac{p_-\cdot J}{k\cdot p_+}\right )
\left (\frac{p_+\cdot J^*}{k\cdot p_-}-
\frac{p_-\cdot J^*}{k\cdot p_+}\right ) \right \}d\Phi \equiv
\frac{e^6}{4(2\pi)^8}~|A_{ISR}|^2d\Phi, \nonumber
\end{eqnarray}
where $d\Phi$ stands for the Lorentz invariant phase space
$$d\Phi=\frac{d\vec{k}}{2\omega}~\frac{d\vec{q}_+}{2E_+}~
\frac{d\vec{q}_-}{2E_-}~
\frac{d\vec{q}_0}{2E_0}~\delta(p_++p_--k-q_+-q_--q_0) $$
and $Q^2=(q_++q_-+q_0)^2=4E(E-\omega)$ is the photon virtuality, $E$ being the
beam energy and $\omega$ -- the energy of the $\gamma$ quantum. Particle
4-momenta assignment can be read from the corresponding diagrams presented
in Fig.\ref{Fig1}.

\noindent The current matrix element $J_\mu$ has a general form
\begin{equation}
J_\mu=\epsilon_{\mu\nu\sigma\tau}q_+^\nu q_-^\sigma q_0^\tau
~F_{3\pi}(q_+,q_-,q_0).
\label{eq2} \end{equation}
For the $F_{3\pi}$ form-factor, which depends only on invariants
constructed from the pions 4-momenta, we will take the expression from \cite{9}
\begin{equation}
F_{3\pi}=\frac{\sqrt{3}}{(2\pi)^2f_\pi^3}\left [\sin{\theta}\cos{\eta}
~R_\omega (Q^2)-\cos{\theta}\sin{\eta}~R_\phi (Q^2)\right ]
\left ( 1-3\alpha_K-\alpha_K H \right ) .
\label{eq3} \end{equation}
Here $\alpha_K\approx 0.5$, $f_\pi\approx 93~\mathrm{MeV}$ is the pion decay
constant, $\eta=\theta-\arcsin{\frac{1}{\sqrt{3}}}\approx 3.4^\circ$
characterizes the departure of the $\omega$-$\phi$ mixing from the ideal one,
and
$$ H=R_\rho(Q_0^2)+R_\rho(Q_+^2)+R_\rho(Q_-^2),$$
where
$$ Q_0^2=(q_++q_-)^2,\;\;
Q_+^2=(q_0+q_+)^2,\;\;Q_-^2=(q_0+q_-)^2. $$
The dimensionless Breit-Wigner factors have the form
$$R_V(Q^2)=\left [ \frac{Q^2}{M_V^2}-1+i\frac{\Gamma_V}{M_V}\right ]^{-1},
\;\; R_\rho(Q^2)=\left [ \frac{Q^2}{M_\rho^2}-1+
i\frac{\sqrt{Q^2}\Gamma_\rho(Q^2)}{M_\rho^2}\right ]^{-1}, $$
where $V=\omega, \phi$ and for the $\rho$ meson the energy dependent width
is used
$$
\Gamma_\rho(Q^2)=\Gamma_\rho \frac{M_\rho^2}{Q^2}\left (\frac{Q^2-4m\pi^2}
{M_\rho^2-4m\pi^2}\right )^{3/2}.
$$
The last term in (\ref{eq1}) is completely irrelevant for VEPP-2M energies
if the hard photon is emitted at a large angle. So we will neglect it in the
following.

\section{Final state radiation}
\begin{figure}[hpb]
  \begin{center}
\mbox{\epsfig{figure=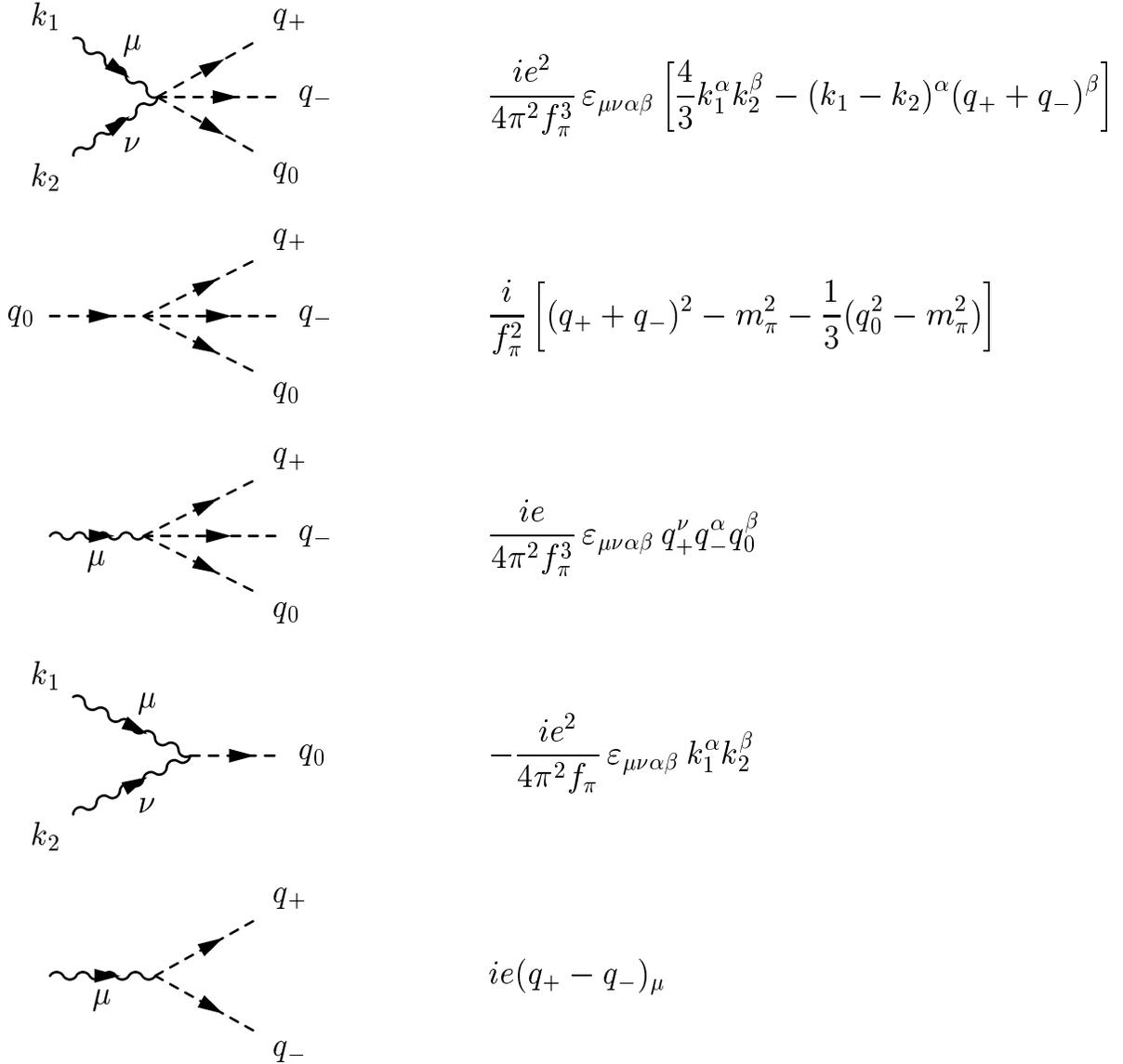
                               ,  height=15.3cm}}
   \end{center}
\caption {Interaction vertexes relevant for final state radiation.}
\label{Fig2}
\end{figure}

To describe final state radiation (FSR), we use the effective low-energy
Wess-Zumino-Witten Lagrangian \cite{10}. The relevant piece of this
Lagran\-gian is reproduced below
\begin{eqnarray} & & \hspace*{30mm}\left . \left .
{\cal{L}}= \frac{f_\pi^2}{4}Sp \right [D_\mu U(D_\mu U)^+
+\chi U^++U\chi^+ \right ]- \nonumber \\ & & \hspace*{-9.2mm}
\left . \left . \left . \left .
-\frac{e}{16\pi^2}\epsilon^{\mu\nu\alpha\beta} A_\mu Sp \right [Q \right \{
(\partial_\nu U)(\partial_\alpha U^+)(\partial_\beta U)U^+-
(\partial_\nu U^+)(\partial_\alpha U)(\partial_\beta U^+)U \right \} \right ]
- \nonumber \\ & & \hspace*{-9mm}
- \frac{ie^2}{8\pi^2}\epsilon^{\mu\nu\alpha\beta}(\partial_\mu A_\nu)
A_\alpha Sp\left [Q^2(\partial_\beta U)U^++Q^2U^+(\partial_\beta U)+
\frac{1}{2}
QUQU^+(\partial_\beta U)U^+ - \right .\nonumber \\ & & \hspace*{30mm} \left .
-\frac{1}{2}QU^+QU(\partial_\beta U^+)U
\right ].   \label{eq4} \end{eqnarray}
Here $U=\exp{\left (i\frac{\sqrt{2}P}{f_\pi}\right )}$, $D_\mu U=\partial_\mu
U+ieA_\mu [Q,U]$, $Q=\mathrm{diag}\left ( \frac{2}{3},-\frac{1}{3},
-\frac{1}{3}\right )$ is the quark charge matrix, and the terms with 
$\chi=B\, \mathrm{diag}\left (m_u,m_d,m_s \right )$
introduce explicit chiral symmetry breaking due to nonzero quark masses. The
constant $B$ has dimension of mass and is determined through the equation
$Bm_q=m_\pi^2,\; m_q=m_u\approx m_d.$ The pseudoscalar meson matrix $P$ has
its standard form
$$ P = \left ( \begin{array}{ccc} \frac{1}{\sqrt{2}}\pi^0+\frac{1}
{\sqrt{6}}\eta & \pi^+
& K^+ \\ \pi^- & -\frac{1}{\sqrt{2}}\pi^0+\frac{1}{\sqrt{6}}\eta & K^0 \\
K^- & \bar K^0 & -\frac{2}{\sqrt{6}}\eta \end{array} \right ). $$
It is straightforward to get from (\ref{eq4}) the relevant interaction
vertices shown in Fig.\ref{Fig2}.

\begin{figure}[hpb]
  \begin{center}
\mbox{\epsfig{figure=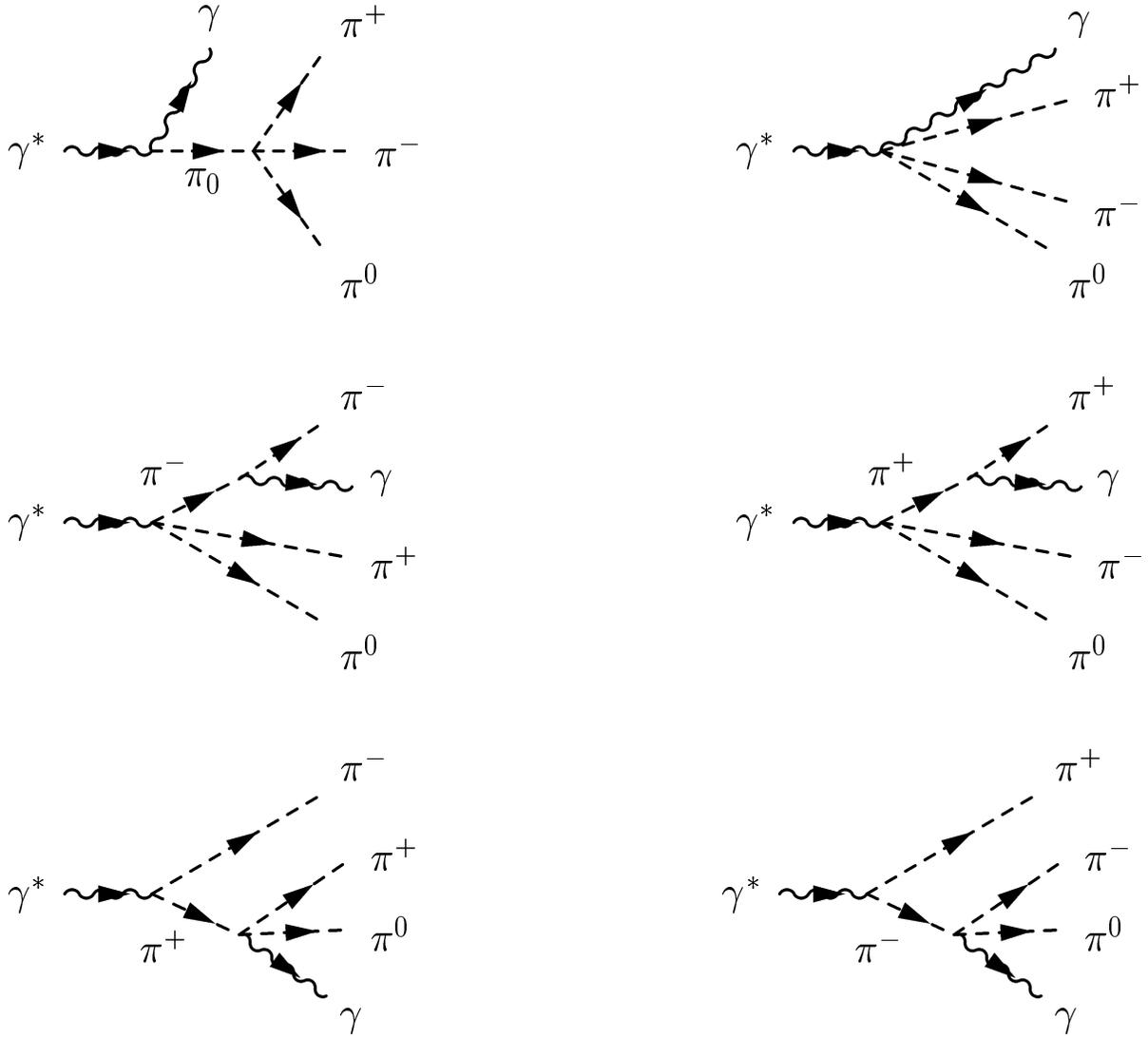
                               ,  height=14.5cm}}
   \end{center}
\caption { $\gamma^*\to\pi^+\pi^-\pi^0\gamma$ transition diagrams.}
\label{Fig3}
\end{figure}

\noindent Using these Feynman rules, one can calculate the $\gamma^*\to
\pi^+\pi^-\pi^0\gamma$ amplitude originated from the diagrams shown in
Fig.\ref{Fig3}.

\noindent The result is
\begin{equation}
A_{\mu\nu}(\gamma_\mu^*\to 3\pi\gamma_\nu)=\frac{ie^2}{4\pi^2f_\pi^3}
T_{\mu\nu},
\label{eq5} \end{equation}
\noindent where ($Q=q_++q_-+q_0+k$ is the virtual photon 4-momentum)
\begin{eqnarray} & &
T_{\mu\nu}=\epsilon_{\mu\nu\alpha\beta}Q^\alpha k^\beta \left ( 1-
\frac{(q_++q_-)^2-m_\pi^2}{(Q-k)^2-m_\pi^2}\right )+
\epsilon_{\mu\nu\alpha\beta}(Q+k)^\alpha q_0^\beta + \nonumber \\ & &
+\epsilon_{\mu\lambda\alpha\beta}Q^\alpha q_0^\beta \left (
\frac{(2q_-+k)_\nu q_+^\lambda}{2q_-\cdot k}+
\frac{(2q_++k)_\nu q_-^\lambda}{2q_+\cdot k}\right )-  \label{eq6} \\ & &
-\epsilon_{\nu\lambda\alpha\beta}
k^\alpha q_0^\beta \left (\frac{(2q_--Q)_\mu q_+^\lambda}{Q^2-2q_-\cdot Q}+
\frac{(2q_+-Q)_\mu q_-^\lambda}{Q^2-2q_+\cdot Q} \right ). \nonumber
\end{eqnarray}
The tensor $T_{\mu\nu}$ is gauge invariant
$$Q^\mu T_{\mu\nu}=k^\nu T_{\mu\nu}=0.$$

\noindent Note that our result for $A_{\mu\nu}(\gamma_\mu^*\to 3\pi
\gamma_\nu)$ is in agreement with the known result \cite{11,12} for the
$\gamma^*\gamma^* \to 3\pi$ amplitude (these two amplitudes are connected by
crossing symmetry, of course).

 If $J_\mu^{(\gamma)}$ is the amplitude of the transition
$\gamma^*_\mu\to\pi^+\pi^-\pi^0\gamma$, then the
final state radiation (FSR) contribution to the $e^+e^-\to\gamma^*\to
\pi^+\pi^-\pi^0\gamma$ process cross section is given by \cite{8}
$$ d\sigma_{FSR}=\frac{e^2}{(2\pi)^8~64 E^4}\sum
\limits_\epsilon \left \{\frac{Re(p_+\cdot J^{(\gamma)}~p_-\cdot
J^{(\gamma)*})}{E^2}-J^{(\gamma)}\cdot J^{(\gamma)*} \right \}~
d\Phi\approx $$ $$
\approx \frac{e^2}{(2\pi)^8~ 64 E^4}\sum\limits_\epsilon \left [
|J_1^{(\gamma)}|^2+|J_2^{(\gamma)}|^2 \right ]~ d\Phi,$$
where the sum is over the photon polarization $\epsilon$ and the$z$ axis was
assumed to be along $\vec{p}_-$, but $J_\mu^{(\gamma)}=\epsilon^\nu
A_{\mu\nu}(\gamma^*\to 3\pi\gamma)$. So we can perform the
polarization sum by using $\sum\limits_\epsilon \epsilon_\mu \epsilon_\nu^*=
-g_{\mu\nu}$. By introducing gauge invariant real 4-vectors $t_1$ and $t_2$
via $t_{1\mu}=T_{1\mu},\; t_{2\mu}=T_{2\mu}$, the result can be casted in the
form (note that the norm of the gauge invariant 4-vector is always negative)
\begin{equation}
d\sigma_{FSR}(e^+e^-\to 3\pi\gamma)=\frac{e^6}{(2\pi)^8~ 64 E^4}~\frac{1}
{(2\pi)^4f_\pi^6} \left [ -t_1\cdot t_1 -t_2\cdot t_2\right ]~d\Phi.
\label{eq7} \end{equation}
However, for the photon virtualities of real experimental interest vector meson
effects can no longer be neglected. So we replace (\ref{eq7}) by
$$
d\sigma_{FSR}(e^+e^-\to 3\pi\gamma)=\frac{e^6}{(2\pi)^8~ 64 E^4}~\frac{1}
{(2\pi)^4f_\pi^6} \left [ -t_1\cdot t_1 -t_2\cdot t_2\right ]~K_{BW}
~d\Phi\equiv $$
\begin{equation} \equiv \frac{e^6}{4(2\pi)^8~}~|A_{FSR}|^2 d\Phi,
\label{eq8} \end{equation}
where we have introduced the phenomenological Breit-Wigner factor
$$K_{BW}=3\left |\sin{\theta}\cos{\eta}R_\omega(4E^2)-
\cos{\theta}\sin{\eta}R_\phi(4E^2)\right |^2. $$
This factor is similar to one presented in ISR ( see (\ref{eq3}) ) and tends
to unity as $E\to 0$. It gives about an order of magnitude increase in
$\sigma_{FSR}$ for energies $2E=0.65\div 0.7~\mathrm{GeV}$.

\section{Monte-Carlo event generator}
Although what follows can be considered as a textbook material \cite{13},
 we will nevertheless give a somewhat detailed description of the Monte Carlo
algorithm for the sake of convenience.

The important first step is the following transformation of the Lorentz
invariant phase space. Let $R_n(p^2;m_1^2,\ldots,m_n^2)$ be $n$-particle
phase space
$$R_n(p^2;m_1^2,\ldots,m_n^2)=\int\prod\limits_{i=1}^n \frac{d\vec{q}_i}
{2E_i}~\delta (p-\sum\limits_{i=1}^n q_i). $$
Inserting the identity
$$1=\int d k_1~d\mu_1^2~\delta (p-q_1-k_1)\delta(k_1^2-\mu_1^2)$$
we get
$$R_4(p^2;m_1^2,m_2^2,m_3^2,m_4^2)=\int \frac{d\vec{q}_1}{2E_1}
R_3((p-q_1)^2;m_2^2,m_3^2,m_4^2)=$$ $$= \int \frac{d\vec{q}_1}{2E_1}
dk_1~d\mu_1^2 R_3(k_1^2;m_2^2,m_3^2,m_4^2)\delta (p-q_1-k_1)
\delta(k_1^2-\mu_1^2).$$
However, (note that $(p-q_1)_0=E_2+E_3+E_4>0$)
$$\int \frac{d\vec{q}_1}{2E_1}dk_1\delta(k_1^2-\mu_1^2)\delta (p-q_1-k_1)=
\int \frac{d\vec{q}_1}{2E_1} \frac{d\vec{k}_1}{2k_{10}}
\delta (p-q_1-k_1)=R_2(p^2;m_1^2,\mu_1^2).$$
Therefore,
\begin{equation}
R_4(p^2;m_1^2,m_2^2,m_3^2,m_4^2)=\int d\mu_1^2
R_3(\mu_1^2;m_2^2,m_3^2,m_4^2)R_2(p^2;m_1^2,\mu_1^2).
\label{eq9}\end{equation}
But \cite{13}
$$R_2(p^2;m_1^2,\mu_1^2)=\int\frac{\lambda^{1/2}(p^2;m_1^2,\mu_1^2)}{8p^2}
d\Omega_1^* $$
where $\lambda$ stands for the triangle function and $\Omega_1^*$ describes
the orientation of the $\vec{q}_1$ vector in the $p$-particle rest frame.

It is more convenient to integrate over $q$-particle energy $E^*$ instead
of mass $\mu$, the two being interconnected by the relation $\mu^2=p^2+q^2-
2\sqrt{p^2}E^*$ in the $p$-particle rest frame.

Using the relation \cite{13}
$$\frac{\lambda^{1/2}(p^2;m^2,\mu^2)}{2\sqrt{p^2}}=\mu
\sqrt{{\bar \gamma}^2-1}, $$
where $\bar \gamma$ is the $\gamma$-factor of the ``particle'' (subsystem)
with the invariant mass $\mu$, after repeatedly using (\ref{eq9})  we get
$$ R_4=\int\frac{1}{2}\sqrt{{\bar \gamma}_1^2-1}~dE_1^*d\Omega_1^*
\frac{1}{2}\mu_1\sqrt{{\bar \gamma}_2^2-1}dE_2^*d\Omega_2^*\frac{1}{2}|\vec{
p}\,_3^*|d\Omega_3^*, $$
where $\vec{p}\,_3^*$ momentum is in the rest frame of the (3,4) subsystem, and
$E_2^*,$  $\Omega_2^*,$ $\;{\bar \gamma}_2$ are in the rest frame of the 
(2,3,4) subsystem.

Now it is straightforward to rewrite the differential cross-section
in the following form:
\begin{equation}
d\sigma(e^+e^-\to 3\pi\gamma)=\frac{\alpha^3}{2\pi^2}|A|^2f d\Phi^*,
\label{eq10} \end{equation}
where $|A|^2= |A_{ISR}|^2+|A_{FSR}|^2$ (we do not take into account
the interference between the initial and final state radiations. This 
interference
integrates to zero if we do not distinguish between negative and positive
$\pi$-mesons),
\begin{equation}
f=\mu_1 (\omega_{max}-\omega_{min})
(E_{0\, max}^*-E_{0\, min}^*)
\sqrt{(E_-^{*2}-m_\pi^2)({\bar \gamma}_1^2-1)({\bar \gamma}_2^2-1)},
\label{eq11} \end{equation}
and
\begin{equation}
d\Phi^*=\frac{d\omega}{(\omega_{max}-\omega_{min})}~\frac{d\varphi}{2\pi}
\frac{
d\cos{\theta}}{2}~\frac{dE_0^*}{(E_{0\, max}^*-E_{0\, min}^*)}
\frac{d\varphi_0^*}{2\pi}\frac{d\cos{\theta_0^*}}{2}
\frac{d\varphi_-^*}{2\pi}\frac{d\cos{\theta_-^*}}{2}.
\label{eq12} \end{equation}
The upper and lower limits for energies are
$$\omega_{max}=\frac{s-9m_\pi^2}{2\sqrt{s}},\;\;
E_{0\, max}^*=\frac{\mu_1^2-3m_\pi^2}{2\mu_1},\;\; E_{0\, min}^*=m_\pi.$$
The minimal photon energy $\omega_{min}$ is an external experimental cut.
At last, $|A_{ISR}|^2$ and $|A_{FSR}|^2$ can be read from the corresponding
expressions (\ref{eq1}) and (\ref{eq8}), respectively.

According to (\ref{eq10}), we can generate $e^+e^-\to\pi^+\pi^-\pi^0\gamma$
events in the cms frame by  the following algorithm.

$\bullet$
generate the photon energy $\omega$ as a random number uniformly distributed
from $\omega_{min}$ to $\omega_{max}$. Calculate for the
$S_1=(\pi^+\pi^-\pi^0)$ subsystem the energy ${\bar E}_1=2E-\omega$, invariant
mass ${\bar \mu}_1=\sqrt{4E(E-\omega)}$ and the Lorentz factor 
${\bar \gamma}_1={\bar E}_1/{\bar \mu}_1$.

$\bullet$
generate a random number ${\bar\varphi}_1$ uniformly distributed
in the interval $[0,2\pi]$ and take it as an azimuthal angle of the $S_1$
subsystem velocity vector in the cms frame. Generate another uniform random
number in the interval $[-\cos{\theta_{min}},\cos{\theta_{min}}]$ and take it
as a $\cos{{\bar \theta}_1},\;{\bar \theta}_1$ being the
polar angle of the $S_1$ subsystem velocity vector in the cms frame. This
defines the unit vector
$\vec{n}_1=(\sin{{\bar \theta}_1}\cos{{\bar \varphi}_1},
\sin{{\bar \theta}_1}\sin{{\bar \varphi}_1},
\cos{{\bar \theta}_1})$ along the $S_1$ subsystem  velocity; $\theta_{min}$ is
the minimal photon radiation angle -- an external experimental cut.

$\bullet$ construct the photon momentum in the cms frame $\vec{k}=
-\omega \vec{n}_1$.

$\bullet$ generate the $\pi^0$-meson energy $E_0^*$ in the $S_1$ rest
frame as a random number uniformly distributed from
$E_{0\, min}^*$ to $E_{0\, max}^*$. Calculate for the $S_2=(\pi^+,\pi^-)$
subsystem the energy ${\bar E}_2={\bar \mu}_1-E_0^*$, invariant mass
${\bar \mu}_2=\sqrt{{\bar \mu}_1^2+m_\pi^2-2{\bar \mu}_1E_0^*}$ and the Lorentz
factor ${\bar \gamma}_2={\bar E}_2/{\bar \mu}_2$.

$\bullet$
generate a random number ${\bar\varphi}_2$ uniformly distributed
in the interval $[0,2\pi]$ and take it as an azimuthal angle of the $S_2$
subsystem velocity vector in the $S_1$ rest frame. Generate another uniform
random number in the interval $[-1,1]$ and
take it as a $\cos{{\bar \theta}_2},\;{\bar \theta}_2$ being the
polar angle of the $S_2$ subsystem velocity vector in the $S_1$ rest frame.
This defines the unit vector along the $S_2$ subsystem  velocity in the 
$S_1$ rest frame
$\vec{n}_2=(\sin{{\bar \theta}_2}\cos{{\bar \varphi}_2},
\sin{{\bar \theta}_2}\sin{{\bar \varphi}_2},
\cos{{\bar \theta}_2})$.

$\bullet$
construct $\vec{q}_0^*=-\sqrt{E_0^{*2}-m_\pi^2}~\vec{n}_2$ -- the
$\pi^0$-meson momentum in the $S_1$ rest frame.

$\bullet$
generate $\varphi_-^*$ and $\cos{\theta_-^*}$ in the manner analogous to what
was described above for ${\bar\varphi}_2$ and $\cos{{\bar \theta}_2}$ and
construct the unit vector along the $\pi^-$ meson velocity in the $S_2$ rest
frame $\vec{n}_3=(\sin{\theta_-^*}\cos{\varphi_-^*}, \sin{\theta_-^*}
\sin{\varphi_-^*},\cos{\theta_-^*})$.

$\bullet$
construct the $\pi^-$-meson 4-momentum in the $S_2$ rest frame
$E_-^*={\bar \mu}_2/2$, $\vec{q}_-^*=\sqrt{E_-^{*2}-m_\pi^2}~\vec{n}_3$.

$\bullet$
construct the $\pi^+$-meson 4-momentum in the $S_2$ rest frame
$E_+^*={\bar \mu}_2/2$, $\vec{q}_+^*=-\vec{q}_-^*$.

$\bullet$
transform $\pi^0$-meson 4-momentum from the $S_1$ rest frame back to the
cms frame.

$\bullet$
transform $\pi^-$ and $\pi^+$ mesons 4-momenta firstly from the $S_2$ rest
frame to the $S_1$ rest frame and then back to the cms frame.

$\bullet$
for the generated 4-momenta of the final state particles, calculate $z=|A|^2f$.

$\bullet$
generate a random number $z_R$ uniformly distributed in the interval from 0
to $z_{max}$ where $z_{max}$ is some number majoring $|A|^2f$ for all
final state 4-momenta allowed by 4-momentum conservation.

$\bullet$
if $z\ge z_R$, accept the event that is the generated 4-momenta of the
$\pi^+,\;\pi^-$ and $\pi^0$ mesons and the photon. Otherwise repeat the whole
procedure.

\section{Soft and collinear photon corrections}
We assume that the photon in the $e^+e^-\to 3\pi\gamma$ reaction is hard
enough $\omega>\omega_{min}$ and radiated at large angle $\theta>\theta_{min}$
so that it could be detected by experimental equipment (a detector).
In any process with accelerated charged particles soft photons are emitted
without being detected because a detector has finite energy resolution. Even
moderately hard photons can escape detection in some circumstances. How
important are such effects? Naively every photon emitted brings extra factor
$e$ in the amplitude and so a small correction is expected. But this
argument (as well as the perturbation theory) breaks down for soft photons.
When an electron (positron) emits a soft enough photon, it nearly remains 
on the
mass shell, thus bringing a very large propagator in the amplitude. Formal
application of the perturbation theory gives an infinite answer to the
correction due to soft photon emission because of this pole singularity. It
is well known \cite{14} how to deal with this infrared divergence. In real
experiments very low energy photons never have enough time and space to be
formed, because of a finite size
of the laboratory. So we have a natural low energy cut-off. A remarkable fact,
 however, is that the observable cross sections do not depend on the actual
form of the cut-off because singularities due to real and virtual soft photons
cancel each other \cite{15}. The net effect is that the soft photon
corrections, summed to all orders of perturbation theory, factor out as
some calculable, so called Yennie-Frautschi-Suura exponent \cite{14}.

Collinear radiation of (not necessarily soft) photons by highly relativistic
initial electrons (positrons) is another source of big corrections which
should also be treated non-perturbatively. Unlike soft photons, however, the
matrix element for a radiation of an arbitrary number of collinear photons
is  unknown. Nevertheless, there is a nice method (the so called Structure
Functions method) \cite{16} which enables one to sum leading collinear 
(and soft)
logarithms. The corrected cross-section, when radiation of unnoticed photons
with total energy less than $\Delta E\ll E$ is allowed, looks like
\cite{16}
\begin{equation}
\tilde \sigma (s)=\int\limits_0^{\Delta E}\frac{d\omega}{\omega}~
\sigma(4E(E-\omega))~\beta \left (\frac{\omega}{E}\right )^\beta \left [ 1+
\frac{3}{4}\beta+\frac{\alpha}{\pi}\left (\frac{\pi^2}{3}-\frac{1}{2}\right )
\right ], \label{eq13} \end{equation}
where $\beta=\frac{2\alpha}{\pi}\left (\ln{\frac{s}{m_e^2}}-1\right )$ and
we have omitted some higher order terms.

In our case the hard photon is well separated (because of $\omega>
\omega_{min},\;\theta>\theta_{min}$ cuts) from the soft and collinear regions
of the phase space. So equation (\ref{eq13}) is applicable and it indicates
that the soft and collinear corrections to the cross-section of the process
$e^+e^-\to 3\pi\gamma$ do not exceed 20\% when
$\Delta E\sim\omega_{min}=30~{\mathrm{MeV}},\;\theta_{min}=
20^\circ$ and  $E=0.7~
{\mathrm{GeV}}$. Such corrections are irrelevant for the present VEPP-2M
statistics but may become important in future high statistics experiments.

\section{Numerical results and conclusions}
\begin{figure}[hpb]
  \begin{center}
\mbox{\epsfig{figure=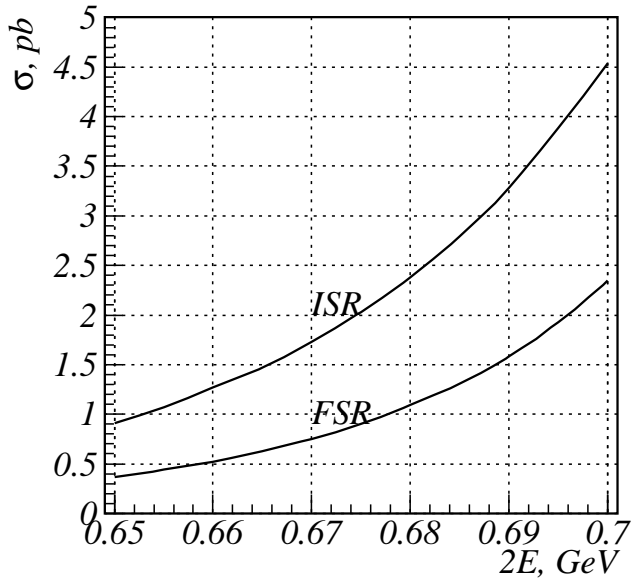
                               ,  height=8.0cm}}
   \end{center}
\caption {ISR and FSR contributions to the $e^+e^-\to 3\pi\gamma$ cross
section. }
\label{Fig4}
\end{figure}

In Fig.\ref{Fig4}, numerical results are shown for $\sigma(e^+e^-\to 3\pi
\gamma)$ with $\omega_{min}=30~\mathrm{MeV},\; \theta_{min}=20^\circ$. As
expected, the cross section is small, only few picobarns, for energies
$0.65\div 0.7~\mathrm{GeV}$.

\begin{figure}[hpb]
\begin{minipage}[t]{0.47\textwidth}
\centerline{\epsfbox{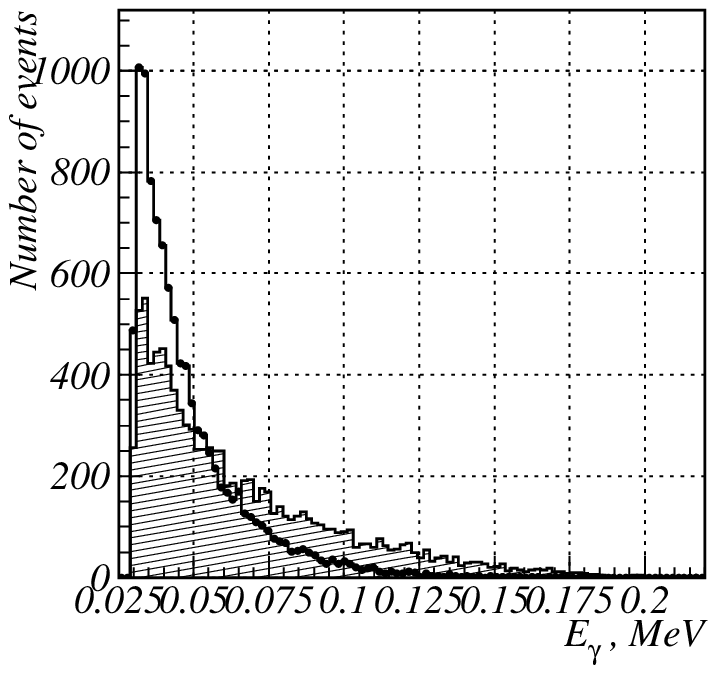}}
\caption {The photon energy distributions for ISR and FSR (hatched
histogram). }
\label{Fig5}
\end{minipage}
\hfill
\begin{minipage}[t]{0.49\textwidth}
\centerline{\epsfbox{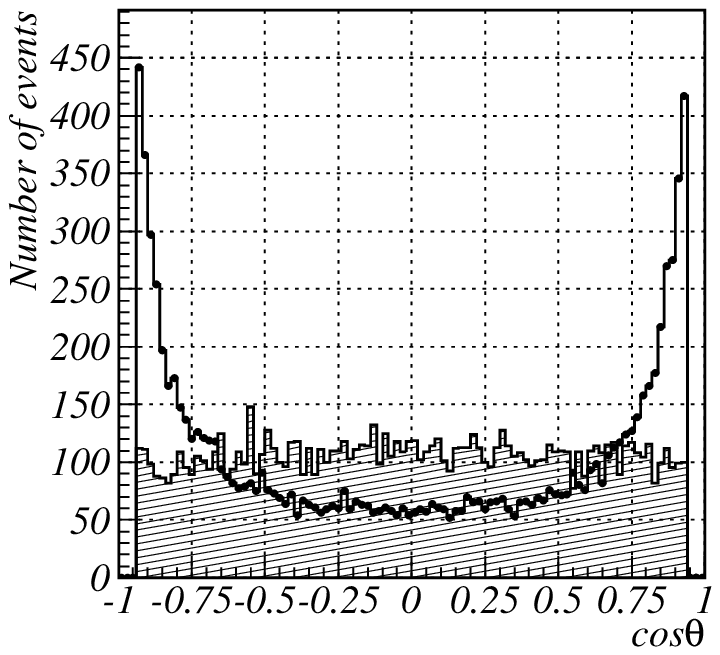}}
\caption {The photon angular distributions for ISR and FSR (hatched
histogram). }
\label{Fig6}
\end{minipage}
\hfill
\end{figure}

\noindent This figure shows also that FSR contributes
significantly at such low energies.
So if future $\phi$-factory experiments produce high enough
statistics in this energy region, the study of FSR will become realistic.
FSR and ISR give different angular and energy distributions for the photon
as illustrated by Fig.\ref{Fig5} and Fig.\ref{Fig6}. This fact can be used
for the FSR separation from a somewhat more intensive ISR.

Let us note, however, that the model considered here is not valid in the
$\phi$-meson region -- very far from the threshold. At that the status of
uncertainties in the ISR and FSR contributions are different. We believe
that the ISR amplitude remains accurate enough even in the $\phi$-meson
region. This belief stems from the fact that all relevant vector meson
effects are already included in the ISR amplitude (\ref{eq3}). The situation
with the FSR amplitude is different. Our phenomenological Breit-Wigner factor
mimics only some part of the vector meson effects. To estimate the
corresponding uncertainty in $\sigma_{FSR}$, let us try some other choices
for $K_{BW}$ which also have the correct low energy limit. If  in the
expression for the $K_{BW}$ factor we make the change
$$R_\omega(4E^2)\to \frac{1}{2} \left [R_\omega(4E^2)+ R_\rho(4E^2)
\right ],$$
$\sigma_{FSR}$ will be lowered by $\sim 5\%$ for $2E=0.65~\mathrm{GeV}$, and
by $\sim 25\%$ for $2E=0.7~\mathrm{GeV}$. In the FSR  amplitude the 
$\rho$-meson contributes via a number of various diagrams. For
example, the $\gamma^*\to\rho\to\rho^+\rho^-\to\pi^+\pi^0\pi^-\gamma$ 
intermediate
state, which has no counterpart in the $\omega$ meson contribution, gives
the following piece of the $T_{\mu\nu}$ tensor
$$T_{\mu\nu}^{(3\rho)}=\left . \left. -\frac{\alpha_K}{2} R_\rho(4E^2)
\right \{A_1-2A_2-2A_3\right \},$$
where
$$A_1=
\epsilon_{\nu\alpha\beta\lambda}~(Q-2q_0)^\alpha k^\beta
\left [ (q_++q_0-q_- -k)_\mu~
q_-^\lambda \,Y_- + \right .$$ $$ \left .+
(q_-+q_0-q_+ -k)_\mu~ q_+^\lambda \,Y_+\right ],$$
$$A_2=\epsilon_{\nu\alpha\beta\lambda}~Q^\alpha k^\beta \left [
(q_+-q_0)_\mu~
q_-^\lambda \,Y_- + (q_--q_0)_\mu~ q_+^\lambda \,Y_+\right ],$$
$$A_3=\epsilon_{\mu\nu\beta\lambda}~k^\beta \left [ Q\cdot(q_+-q_0)~
q_-^\lambda \,Y_- +Q\cdot(q_--q_0)~q_+^\lambda \, Y_+\right ],$$
and
$$Y_\mp=\frac{M_\rho^2}{\left [(q_\mp+k)^2-M_\rho^2\right ]
\left [(q_\pm+q_0)^2-M_\rho^2\right ]}.$$
If we include this contribution, and besides ensure that the remaining part 
of the FSR amplitude (\ref{eq6}) also takes
$$K_{BW}= |R_\rho(4E^2)|^2$$
in the role of the phenomenological Breit-Wigner factor,
the FSR cross section will be lowered by $\sim 10\%$  for $2E=0.65~
\mathrm{GeV}$, and by $\sim 35\%$ for $2E=0.7~\mathrm{GeV}$.

This uncertainty
in the FSR magnitude is irrelevant for the present VEPP-2M statistics.
For future high precision experiments a systematic inclusion of all relevant
vector meson effects in the FSR amplitude is desired.

\section*{Acknowledgments}
One of us (EAK) is grateful to Heisenberg-Landau Fund 2000-02 and to RFFI
Grant 99-02-17730. We are grateful to G. Sandukovskaya for the help.



\begin{thebibliography}{99}
\bibitem{1}
H.~N.~Brown {\it et al.}  [Muon g-2 Collaboration],
Phys.\ Rev.\ Lett.\  {\bf 86}, 2227 (2001).
\bibitem{2} For review see, for example \newline
A.~Czarnecki and W.~J.~Marciano,
Phys.\ Rev.\ D {\bf 64}, 013014 (2001).
\bibitem{3}
K.~Melnikov,
Int.\ J.\ Mod.\ Phys.\ A {\bf 16}, 4591 (2001);
\newline J.~F.~De Troconiz and F.~J.~Yndurain,
hep-ph/0106025;
\newline M.~Knecht and A.~Nyffeler,
hep-ph/0111058.
\bibitem{4} M.~N.~Achasov {\it et al.},
hep-ex/0010077.
\newline  R.~R.~Akhmetshin {\it et al.}  [CMD-2 Collaboration],
Nucl.\ Phys.\ A {\bf 675}, 424C (2000).
\bibitem{5} R.~R.~Akhmetshin {\it et al.}  [CMD-2 Collaboration],
hep-ex/9904027.
\bibitem{6} R.~R.~Akhmetshin {\it et al.}  [CMD-2 Collaboration],
Phys.\ Lett.\ B {\bf 476}, 33 (2000).
\bibitem{7} A.~Aloisio {\it et al.}  [KLOE Collaboration],
hep-ex/0107023.
\bibitem{8} G.~Bonneau and F.~Martin,
Nucl.\ Phys.\ B {\bf 27}, 381 (1971).
\bibitem{9} E.~A.~Kuraev and Z.~K.~Silagadze,
Phys.\ Atom.\ Nucl.\  {\bf 58}, 1589 (1995)
\bibitem{10} E.~Witten,
Nucl.\ Phys.\ B {\bf 223}, 422 (1983).
\newline J.~Wess and B.~Zumino,
Phys.\ Lett.\ B {\bf 37}, 95 (1971).
\bibitem{11} J.~W.~Bos, Y.~C.~Lin and H.~H.~Shih,
Phys.\ Lett.\ B {\bf 337}, 152 (1994).
\bibitem{12} P.~Talavera, L.~Ametller, J.~Bijnens, A.~Bramon and F.~Cornet,
Phys.\ Lett.\ B {\bf 376}, 186 (1996);
\newline L.~Ametller, J.~Kambor, M.~Knecht and P.~Talavera,
Phys.\ Rev.\ D {\bf 60}, 094003 (1999).
\bibitem{13} E.~Byckling and K.~Kajantie, {\it Particle Kinematics}.
Wiley, 1973.
\bibitem{14} D.~R.~Yennie, S.~C.~Frautschi and H.~Suura,
Annals Phys.\  {\bf 13}, 379 (1961).
\bibitem{15} F.~Bloch and A.~Nordsieck,
Phys.\ Rev.\  {\bf 52}, 54 (1937).
\newline D.~R.~Yennie, {\it in} Lectures on strong and electromagnetic
interactions, p. 166 (Brandies Summer Institute in Theoretical Physics,
Waltman, Mass., 1963)
\bibitem{16} E.~A.~Kuraev and V.~S.~Fadin,
Sov.\ J.\ Nucl.\ Phys.\  {\bf 41}, 466 (1985).
\newline For review see
M.~Skrzypek,
Acta Phys.\ Polon.\ B {\bf 23}, 135 (1992).

\end{thebibliography}
\end{document}